\documentclass[12 pt]{article}

\usepackage{amsmath,amstext,amsgen,amsbsy,amsopn,amsfonts,graphicx, overcite}

\begin{document}

\title{Nonadiabatic Dynamics in Semiquantal Physics}

\author{Mason A. Porter \\ \\ Center for Applied Mathematics \\ \\ Cornell 
University}

\date{April, 2001 \\ Revised July, 2001}

\maketitle

\begin{centering}
\section*{Abstract}
\end{centering}

\vspace{.1 in}

	Every physical regime is some sort of approximation of reality.  One 
lesser-known realm that is the semiquantal regime, which may be used to 
describe systems with both classical and quantum subcomponents.  In the 
present review, we discuss nonadiabatic dynamics in the semiquantal regime.  
Our primary concern is electronic-nuclear coupling in polyatomic molecules, 
but we discuss several other situations as well.  We begin our presentation by
 formulating the semiquantal approximation in quantum systems with 
degrees-of-freedom that evolve at different speeds.  We discuss nonadiabatic 
phenomena, focusing on their relation to the Born-Oppenheimer approximation.  
We present several examples--including Jahn-Teller distortion in molecules and
 crystals and the dynamics of solvated electrons, buckyballs, nanotubes, atoms
 in a resonant cavity, SQUIDs, quantum particle-spin systems, and micromasers.
  We also highlight vibrating quantum billiards as a useful abstraction of 
semiquantal dynamics.

\vspace{.3 in}

\subsection*{PAC NOS 03.65.Sq, 02.70.Ns, 05.45.-a, 05.45.Mt}

\begin{centering}
\section{Introduction}
\end{centering}

	Every physical regime approximates reality in some form or another.  
In continuum mechanics, one ignores the fact that a solid or fluid is composed
 of a finite number of discretely spaced particles because it is not necessary
 to consider this at the scale under consideration.  In classical mechanics, 
one does consider discrete objects, but quantities such as energy and light 
are permitted to vary continuously.  Indeed, there are so many photons in this
 regime that one would not notice the ensuing difference in illumination were 
a single one removed.  In quantum physics, these quantities are treated as 
discrete--they have been quantized--and one expresses concepts such as 
position and momentum as operators rather than simply vectors.  This regime is 
an approximation of the even more finely grained domain of quantum field 
theory, which may be in turn an approximation of even more intricate theories.
  Despite this outline, the picture is far from complete, as there are several
 regimes not mentioned above as well as others that lie at the borders between
 the regimes discussed above.  For example, between the fully quantum regime 
and the classical one lie the semiclassical, quasiclassical, and semiquantal 
regimes, which--despite their nomenclature--are not the same.  The 
semiclassical regime is a well-studied physical approximation to quantum 
mechanics.  It is the domain of methods such as the WKB approximation and 
quantum chaology, the study of the quantum signatures of classical 
chaos.\cite{gutz}  One obtaines a semiclassical description from the fully 
quantal theory by taking the well-defined asymptotic limit $\hbar 
\longrightarrow 0$.\cite{sakurai}  

	Less studied than the semiclassical regime is the semiquantal one.  
This latter description of physics has been analyzed far less than the 
semiclassical one in part because nobody has found a completely satisfactory 
asymptotic procedure to pass from the fully quantized regime to the 
semiquantal one.  Nevertheless, there are several situations for which 
semiquantal physics is appropriate.  Such systems are characterized by a 
mixture of classical and quantum physics.  One may obtain a semiquantal 
description, for example, by coupling a classical system to a 
quantum-mechanical one.  Moreover, semiquantal systems arise naturally when 
one uses the adiabatic or Born-Oppenheimer approximation\cite{ezra}, which 
provides a widely accepted procedure for dividing a quantum-mechanical system 
into slow and fast subsystems.  One begins this approximation by quantizing 
the fast subsystem, which consists of the 
\begin{itshape}electronic\end{itshape} degrees-of-freedom (in the language of 
chemical physics).  If one obtains well-separated energy levels, then one may 
also quantize the slow subsystem, which consists of the 
\begin{itshape}nuclear\end{itshape} degrees-of-freedom (which can be either 
vibrational or rotational).  If, however, the electronic eigenenergies of a 
$d$-state system are close to each other, then one ignores the rest of the 
spectrum, thereby obtaining a system described by $d$ electronic energy levels
 (each of which corresponds to the full contribution of a single eigenstate) 
that are coupled to a multitude of nuclear states.  The semiquantal 
approximation consists of modeling these nuclear states as a continuum.  That 
is, we treat the nuclear degrees-of-freedom of the present system as classical
 degrees-of-freedom, thereby obtaining a system with coupled classical and 
quantum components.  

	This breakdown of the Born-Oppenheimer approximation is a hallmark of 
\begin{itshape}nonadiabatic\end{itshape} phenomena, which are important in 
the study of inelastic atomic and molecular collisions as well as in bound 
states of molecular systems.  In particular, the Born-Oppenheimer 
approximation breaks down in exactly this manner for excited electron states 
of polyatomic molecules--often as a result of their symmetries.  The 
near-degeneracy (and sometimes exact degeneracy) of several states is a common
 phenomenon in molecules--especially at higher energies.  Both the energy 
spectrum and intramolecular dynamics can vary substantially from those 
observed during adiabatic behavior.\cite{ezra2}  However, it is not easy to 
incorporate nonadiabatic behavior into simple models of molecular dynamics, in
 which the canonical portrait of nuclear motion is described on a single 
well-defined surface of potential energy near the electronic degeneracy.  
Consequently, it is important to develop a semiquantal description of such 
systems that incorporates essential features of the nonadiabatic coupling.

	The purpose of the present paper is to give an elementary presentation
 of the semiquantal approximation and some systems for which it is relevant.  
We focus on problems in which chaotic behavior can occur.  We include few 
calculations and instead provide references to papers and monographs that 
include them.  We begin our discussion by framing semiquantal physics in the 
context of the Born-Oppenheimer approximation before going into a more 
detailed discussion of nonadiabaticity.  We then discuss polyatomic 
molecules\cite{ezra,ezra2}, crystals\cite{vibe}, solvated 
electrons\cite{solve1,solve2}, carbon nanotubes\cite{tube,full}, and 
buckyballs\cite{bucky,bounce}.  These systems may be abstracted mathematically
 in terms of vibrating quantum billiards,\cite{vibline,sazim} which are 
amenable to a semiquantal description because the boundaries are classical and
 the enclosed particles are quantum-mechanical.\cite{nec}  We also survey 
other systems such as a two-level system interacting with the electromagnetic 
field of a laser cavity\cite{berman}, micromasers\cite{haroche}, quantum 
particle-spin systems\cite{real}, SQUIDs\cite{squid}, and nuclear collective 
motion\cite{bulgac}.

\begin{centering}
\section{The Semiquantal Approximation}
\end{centering}

	At issue in nonadiabatic analysis is the exent to which ``classical 
path'' (that is, functional integration\cite{qft}) descriptions are relevant. 
 In other words, one must consider how reasonable is it to treat the nuclear 
(slow, heavy particle) degrees-of-freedom classically in an effective 
potential determined by the quantum dynamics of the electronic (fast, light 
particle) degrees-of-freedom.  If one uses a path integral formulation of 
nonadiabatic scattering amplitudes, one obtains a formal solution to this 
problem.\cite{pech1,pech2}  The exact (semiclassical) effective ``potential'' 
is nonlocal in time, so it must be computed iteratively.  Consequently, 
practical computations of this quantity are almost impossible without 
approximations.  Two ways of dealing with this are the so-called 
``surface-hopping'' approach and the idea of self-consistent matrix 
propagation, which resembles Feynman path integrals in spirit.\cite{ezra,hop}  

	Consider a system with nuclear degrees-of-freedom $Q$ and electron 
degrees-of-freedom $q$.  The quantum dynamics of the full system may be 
expressed in integral form with a propagator\cite{sakurai} (i.e., Green's 
function) $K$ as the kernel of the following integral equation:
\begin{equation}
	\psi(q',Q',t') = \int dq dQ K(q',Q',t'|q,Q,t)\psi(q,Q,t).
\end{equation}
One may equivalently expand $\psi$ in a basis of electronic states 
$\{\varphi_n(q)\}$ to obtain the equation
\begin{equation}
	\psi(q,Q,t) = \sum_n \chi_n(Q,t)\varphi_n(q,t), \label{diab}
\end{equation}
where the dynamics of the nuclear wavefunctions $\chi_n$ are determined using
 the reduced propagator $K_{\beta\alpha}$:
\begin{equation}
	\chi_\beta(Q',t') = \sum_\alpha \int dQK_{\beta\alpha}(Q',t'|Q,t)
\chi_\alpha(Q,t).
\end{equation}
(Note that equation (\ref{diab}) is valid asymptotically (that is, 
adiabatically) only when the electronic and nuclear degrees-of-freedom can be 
separated from each other.\cite{baym,ezra})  The functional 
$K_{\beta\alpha}(Q',t'|Q,t)$ gives the probability amplitude for the quantum 
system to go from state $\alpha$ to state $\beta$ as the nuclear variables 
move from $Q(t)$ to $Q'(t')$.  The probability of being in the electronic 
basis state $\beta$ at time $t'$ is thus given by
\begin{equation}
	P_\beta(t') = \int dQ' |\chi_\beta(Q',t')|^2,
\end{equation}
where $\sum_\beta P_\beta(t') = 1$ by conservation of probability.  One may 
write the Green's function $K_\beta$ as a Feynman path integral, thereby 
expressing it as an integral over all paths $\bar{Q}\left(\bar{t}\right)$ 
connecting the two endpoints $Q(t)$ and $Q'(t')$.  By assuming that the paths 
of the nuclear coordinates with the biggest contribution are those of 
stationary phase $\tilde{R}(t)$, one obtains a semiquantal approximation of 
the reduced propagator.  One integrates only over ``classical'' (stationary 
phase) paths rather than over every path and obtains equations of motion that 
must be solved iteratively (due to temporal nonlocality).  Indeed, the force 
on the trajectory at time $\bar{t}$ depends on both the forward and backward 
propagated wavefunctions, which can only be determined if one knows the full 
trajectory $\tilde{R}(t)$.  One begins the iterative procedure by guessing a 
trajectory, which specifies the electronic component of the Hamiltonian.  One 
then integrates the time-dependent Schr\"odinger equation both forwards and 
backwards from the appropriate boundary states and uses mixed-state solutions 
to determine the force on the trajectory.  

	If the electronic state $\alpha$ is initially occupied, the initial 
wave vector at the outset is given by the components
\begin{gather}
	\chi_i(Q(t),t) = 0 \hspace{.1 in} \forall \hspace{.1 in} i \neq 
\alpha, \notag \\ 
	\chi_\alpha(Q(t),t) = f(Q(t)),
\end{gather}
where $f(Q(t))$ is the shape of the initial nuclear wavefunction on the 
electronic surface $\alpha$.  Hence, the components of the wavefunction at 
time $t'$ are given by
\begin{equation}
	\chi_\beta(Q',t') = \int dQ K_{\beta\alpha}(Q',t'|Q,t)f(Q).
\end{equation}
One subsequently uses a semiquantal approximation of the propagator 
$K_{\beta\alpha}$ in order to obtain the semiquantal expression for the 
advanced nuclear wavefunction at $Q'$ moving over the electronic state 
$\beta$.  In principle, one can obtain the stationary phase paths using a root
 search, as they are specified in terms of a boundary value problem.  One
 considers initial velocities $d\tilde{Q}/dt$ and proceeds iteratively until 
one has found all convergent paths that reach the desired endpoint $Q'$.  
Finally, one computes the wavefunction by integrating over all these path 
contributions from each initial point.

	The surface-hopping method has been applied to scattering problems, 
for which nonadiabatic effects are usually localized.  One assumes that a 
trajectory evolves on a single manifold of adiabatic potential surfaces for 
every nuclear configuration except those near electronic degeneracies.  One 
then  calculates the probability that the trajectory jumps to a nearby 
surface, on which the evolution proceeds adiabatically.  Many systems, 
however, are constantly in regions of near-degeneracy, so one requires a 
dynamical description of nonadiabatic evolution.  It is desirable, moreover, to
 have a scheme with which to analyze the nonadiabatic behavior of bound and 
quasibound molecular states.  If the nuclei in such states are localized in 
regions of electronic degeneracy (or near-degeneracy), one may use 
effective-path methods that couple classical nuclear motion self-consistently 
with quantum electronic motion.\cite{ezra2}  One may assume that the classical
 nuclear motion is determined by its interaction with the electronic system in
 a self-consistent manner.  Using the vibrational and rotational 
(\begin{itshape}rovibrational\end{itshape}) coupling terms in the molecular 
Hamiltonian, one obtains a time-dependent electronic Hamiltonian, which causes
 transitions in the molecule's electronic states because of its dependence on 
the nuclear degrees-of-freedom.  Time-dependence in these electronic states 
leads to a time-dependent nuclear potential, because the molecular Hamiltonian
 depends on the nuclear coordinates.  Simple examples of this sort of 
self-consistent coupling may be abstracted and studied as a vibrating quantum 
billiards, in which the enclosed particle (fast subsystem) is coupled to the 
surrounding wall (slow subsystem).\cite{sazim,ksu}  This abstraction is very 
useful, as it is easily generalized and may also be applied to the study of 
systems in chemical physics (such as polyatomic molecules, solvated electrons,
 Jahn-Teller distortions, and chemical nanostructures).

	We now continue our discussion of path integral methods.  Consider
 a system with one nuclear degree-of-freedom that includes a localized 
crossing of two adiabatic potentials.  It is requisite that the two potentials
 have similar slopes in the crossing region and that the nuclear kinetic 
energy is large compared to the difference between the two potentials in the 
interaction region.\cite{ezra}  Moreover, one needs the electronic states to 
be close enough to each other energetically so that the nuclear 
degrees-of-freedom may be approximated as a continuum.  The nuclear mass is 
assumed to be large relative to the electronic mass, so the nuclear energy 
levels are more finely grained than the electronic eigenenergies.  If the 
nuclear energy levels are sufficiently close together, they are 
well-approximated by a continuum.  The separation of these energy levels 
becomes smaller both as a result of larger nuclear masses and as a consequence
 of closeness of electronic energy levels.  Hence, one requires some 
combination of sufficiently large nuclear mass and sufficiently degenerate 
electronic energy levels in order to approximate the nuclear 
degrees-of-freedom as classical.

	Although this type of self-consistent coupling of classical and 
quantum dynamics has appeared often in the chemical physics literature, there 
remain conceptual difficulties and inconsistencies in the semiquantal 
approximation.  Consider a system that is asymptotically in a two-state 
electronic superposition.  The nuclei undergo some sort of averaged dynamical 
motion that does not correspond to that determined by either of the two 
adiabatic surfaces, despite the fact that one could argue by physical 
reasoning that the latter is the expected behavior.  A way to surmount this 
difficulty is the \begin{itshape}classical electron picture\end{itshape}, 
which facilitates treatment of resonant processes such as 
electronic-vibrational and electronic-rotational energy transfer.\cite{meyer} 
 This method has been applied successfully to several systems describing 
nonadiabatic collisions, including charge transfer in $Na + I$ collisions, the
 quenching of the fluorine atom $F^*(^2 P_\frac{1}{2})$ via collisions with 
$H^+$ or $Xe$, and collinear and three-dimensional systems.  Fully quantal 
calculations are available for some of these systems, and the semiquantal 
analyses produce cross sections and transition probabilities that are 
consistent with these studies.  Additionally, the semiquantal calculations 
provide a correct description of resonant features.\cite{ezra}  

\begin{centering}
\section{The Born-Oppenheimer Approximation and Nonadiabatic Phenomena}
\end{centering}

	It is more difficult to find electron orbits in molecules than in 
atoms because the effective potential felt by the electrons is no longer 
well-approximated as spherically symmetric.  One pictures the molecular 
nucleus as having classical equilibrium positions about which it slowly 
oscillates.  The electrons travel rapidly around the nucleus and are affected 
by the oscillations of the latter.  This perspective is effective because a 
nucleus (with mass $M$) is much more massive than electrons (each of which 
have mass $m$).  The mass ratio $m/M$ is typically about
\begin{equation}
	\frac{m}{M} \approx 10^{-5} \text{ or } 10^{-4},
\end{equation}
so the magnitude of the zero-point motion of the nucleus is far smaller than 
that of the electrons.  (Zero-point motion describes the minimal motion due to 
Heisenberg's Uncertainty Principle.)

	From the perspective of an electron, the nucleus is practically 
stationary.  As long as the electronic energy levels are sufficiently far 
apart, the only effect of the slow nuclear vibrations is to adiabatically 
deform the electronic eigenstates.  A molecule with typical radius $a$ has 
electrons with approximate momenta $\hbar/a$, so the energetic spacing of
 these electrons is about $\hbar^2/m a^2$.  

	From the nuclear point of view, the electrons are a blurry cloud.  The
 electronic wavefunctions distort as the nuclei move, thereby causing small 
changes in the electronic energies.  Additionally, the nuclei tend to move 
towards positions of minimum electronic energy, as if they were immersed in an
 elastic medium formed of electrons.\cite{baym}  Molecular nuclei thus 
oscillate about energy minima, a phenomenon captured by the vibrating quantum 
billiard model of electronic-nuclear coupling.\cite{sazim}  One can estimate 
the frequency $\omega$ of nuclear oscillations by assuming that the nucleus 
resides in a harmonic potential $M\omega^2r^2/2$, where $r$ is the 
displacement of the nucleus from equilibrium.  If this displacement is given 
by the distance $a$, then the electronic energy experiences a change of about 
$\hbar^2/2m a^2$.  As a rough approximation,
\begin{equation}
	\frac{M\omega^2a^2}{2} \approx \frac{\hbar^2}{2m a^2},
\end{equation}
so the nuclear frequency is given by\cite{baym}
\begin{equation}
	\omega \approx \sqrt{\frac{m}{M}}\frac{\hbar}{m a^2}.
\end{equation}
The nuclear vibration energies $\hbar \omega$ are consequently a factor of 
$\sqrt{m/M}$ smaller than the electronic excitation energies and are on the 
order of tenths or hundredths the size of an electron volt.

	The zero-point nuclear energy in a harmonic potential is 
\begin{equation}
	\frac{P^2}{2M} \approx \frac{\hbar \omega}{2},
\end{equation}
so its corresponding zero-point momentum is
\begin{equation}
	P \approx \left(\frac{M}{m}\right)^\frac{1}{4} \frac{\hbar}{a},
\end{equation}
which is about ten times larger than the momentum of an electron.  A typical 
nuclear velocity is thus
\begin{equation}
	v_N = \frac{P}{M} \approx \left(\frac{m}{M}\right)^\frac{3}{4} 
\frac{\hbar}{m a}.
\end{equation}
The nuclear deviation from equilibrium $\delta$ satisfies
\begin{equation}
	\frac{M \omega^2 \delta^2}{2} \approx \frac{\hbar \omega}{2},
\end{equation}
so
\begin{equation}
	\left(\frac{\delta}{a}\right)^2 \approx \frac{\hbar\omega}
{M \omega^2 a^2} \approx \frac{E_N}{E_e} \approx \sqrt{\frac{m}{M}},
\end{equation}
which implies that\cite{baym}
\begin{equation}
	\left(\frac{\delta}{a}\right) \approx \left(\frac{m}
{M}\right)^\frac{1}{4} \approx \frac{1}{10}.
\end{equation}

	In addition to vibrations, one may consider the rotation of the entire
 molecule about its center of mass, although the energy due to such 
excitations is very small since the molecule does not experience much 
distortion as a result of this motion.  If the angular momentum of the 
rotational motion is $\hbar l$, then its accompanying energy is
\begin{equation}
	E_{rot} \approx \frac{\hbar^2 l(l+1)}{2 M a^2} \approx \frac{m}{M}E_e.
\end{equation}
In general, a molecular excited state can be decomposed into electronic, 
vibrational, and rotational excitations.  Two examples of vibrational motion 
are pulsing (as in vibrating quantum billiards\cite{nec}) and ``bouncing'' of 
the center-of-mass (as has been proposed as a mechanism for energy transfer in
 buckyballs\cite{bucky}).  Together, the vibrational and rotational 
excitations comprise the nuclear (or 
\begin{itshape}rovibrational\end{itshape}) contribution to the energy.  The 
total energy is given by the sum of the contributions from its three 
components:
\begin{equation}
	E = E_e + E_N + E_{rot}.
\end{equation}

	Let us formalize the preceeding discussion (which is based on the 
presentation of Baym\cite{baym}).  We supplement the above analysis by 
applying the Born-Oppenheimer approximation to the Schr\"odinger equation.  
This scheme provides a widely accepted procedure for dividing quantum systems 
into slow and fast subsystems.  The first step in this approximation is to 
quantize the fast (electronic) subsystem.  If this results in energy levels 
with sufficient separation (because the electronic energy levels are 
sufficiently far apart for the given nuclear and electronic masses), then one 
can also quantize the slow (nuclear) subsystem in order to perform a fully 
quantum analysis.  This results in a familiar spectrum describing the coupling
 between the electronic and nuclear subsystems.  (In the language of physical 
chemistry, we think of the fast system as describing particles such as 
electrons and the slow system as describing nuclear variables.)  If, however, 
the electronic energy levels are sufficiently close together, the situation is
 more complicated.  One uses a semiquantal approximation by modeling as a 
continuum the large number of vibrational states that are coupled to the $d$ 
electronic states.  Such systems are $(d + s)$ degree-of-freedom 
(\begin{itshape}dof\end{itshape}) Hamiltonian systems, where $d$ of the 
\begin{itshape}dof\end{itshape} are quantum-mechanical and the other $s$ are 
classical.  

	The number of classical degrees-of-freedom of such molecular systems 
(in other words, the number of nuclear \begin{itshape}of\end{itshape}) is 
known as the \begin{itshape}degree-of-vibration\end{itshape} (dov) of the 
system.\cite{sazim}  In semiquantal systems, many of which may be abstracted 
mathematically as vibrating quantum billiards, one often observes a form of 
quantum chaos known as \begin{itshape}semiquantum chaos\end{itshape}, where 
the nomenclature reflects the fact that it occurs in the semiquantal 
regime.\cite{ksu,nec,rect}  Another form of quantum chaos, called 
\begin{itshape}quantized chaos\end{itshape} or \begin{itshape}quantum 
chaology\end{itshape}, is studied in the semiclassical and high quantum-number
 regimes.\cite{gutz,atomic}  This latter behavior may be observed by 
fully quantizing the motion of the molecular systems we have been describing. 
 Part of the value of the semiquantal setting is that one may observe chaos 
even in low energy systems, such as nuclei that have been coupled to two-level 
electronic systems consisting of the ground state and the first excited state 
of appropriate symmetry.  In the setting of quantum chaology, one observes 
chaos only in states with high energy.\cite{gutz}  In other words, one must 
pass to the semiclassical or high quantum-number limits in order to observe 
chaotic behavior.  The ground state is not encompassed by these limits, so the
 semiquantal regime is important for capturing chaotic dynamics of low-energy 
states.  Such behavior has been observed experimentally.\cite{ezra}

	Let us now consider the relation of the Born-Oppenheimer approximation
 to nonadiabatic phenomena.  In so doing, we largely follow the presentation 
of Whetten, Ezra, and Grant.\cite{ezra}  They use the so-called 
\begin{itshape}effective path\end{itshape} method, in which the electronic 
degrees-of-freedom are treated in the same manner as the nuclear 
degrees-of-freedom.  One analyzes the nuclear motion classically to derive an 
effective Hamiltonian describing the evolution of the electronic states.  An 
abstraction of this type of system is a vibrating quantum 
billiard\cite{sazim}, in which the one derives evolution equations that may be
 treated as a classical Hamiltonian system.  For a symmetric two-level system 
in that context, the dynamics of the quantum variables occur on the Bloch 
sphere.\cite{bloch}  (Alternatively, one can describe the quantum dynamics of 
the system using action-angle variables.)  After ignoring the quantum setting 
during numerical simulations, one must later interpret one's results in this 
context.  This reinterpretation leads to interesting mathematical and physical
 phenomena.  For example, bound states of molecules may exhibit nonadiabatic 
behavior.  If the nuclei are localized near a degeneracy, effective path 
methods that couple classical nuclear motion with quantum-mechanical 
electronic motion provide a self-consistent (though possibly approximate) 
treatment of the effects of nonadiabatic coupling.  One situation to which 
this has been applied is the Jahn-Teller E $\times$ e system, in which a 
doubly degenerate electronic state is coupled to a doubly degenerate 
vibrational mode.\cite{ezra2}

	Important manifestations of nonadiabatic behavior may be observed in 
simple examples of symmetry-based electronic degeneracy.  Without such 
degeneracy, one may approximate the molecular wavefunction using the 
Born-Oppenheimer (adiabatic) approximation.  Using this scheme, the 
wavefunction is expressed as a product of electronic and nuclear wavefunction.
  One expands the wave in a $d$-dimensional electronic basis when one is near 
a $d$-fold degeneracy.  Such degeneracies are common in the space spanned by 
nuclear (vibrational) coordinates.\cite{ezra}  

	The stationary and spinless Schr\"odinger equation for a single 
molecule is
\begin{equation}
	\left[T_N + H_e(q,Q)\right]\psi_d(q,Q) = E_d\psi_d(q,Q),
\end{equation}
where $T_N$ is the nuclear kinetic energy operator and
\begin{equation}
	H_e(q,Q) \equiv T_e + U_{ee} + U_{eN} + U_{NN} + V
\end{equation}
is the electronic Hamiltonian.  Because of the coupling, $H_e$ depends 
(continuously) on the nuclear coordinates $Q$.  Its components are the 
particle (electronic) kinetic energy $T_e$, the interelectron repulsion 
potential $U_{ee}$, the electron-nuclear attraction $U_{eN}$, the internuclear
 repulsion $U_{NN}$, and an external potential $V$.  The nuclear kinetic 
energy $T_N$ is proportional to $1/M$, so it is a small term in the total 
Hamiltonian.  The Born-Oppenheimer scheme is to calculate the eigenenergies 
and eigenstates of the total molecular Hamiltonian by treating $T_N$ as a 
small perturbation whose expansion parameter is $(m/M)^\frac{1}{4}$, the ratio
 of nuclear vibrational displacement to the spacing between nuclei.

	The molecular Hamiltonian $H$, given by
\begin{equation}
	H = T_N + H_e,
\end{equation}
is the sum of its nuclear and electronic components.  In the vibrating quantum
 billiard model\cite{sazim}, an abstract example of a semiquantal system, the 
nuclear kinetic energy is simply the kinetic energy of the billiard boundary:
\begin{equation}
	T_N \equiv \frac{P^2}{2M},
\end{equation}
where $P$ is the momentum of the boundary and $M$ is its mass.  The only
 electronic Hamiltonians that have been considered in this abstract situation 
are ones without interelectron repulsion, electron-nuclear attraction, and 
internuclear repulsion.  In other words, the particle's electronic 
Hamiltonian $H_e$ is given by the sum of its kinetic energy component $T_e$ 
and the external potential $V$:
\begin{equation}
	H_e \equiv T_e + V,
\end{equation}
where
\begin{equation}
	T_e \equiv K = -\frac{\hbar^2}{2m}\nabla^2,
\end{equation}
and $m$ is the mass of the confined particle.  One would add an interelectron 
repulsion potential when considering a billiard with more than one enclosed 
particle.  Similarly, one would add an electron-nuclear attraction term if 
considering vibrating quantum billiards with rebound.  

	The particle confined within the billiard is constrained to collide 
elastically against the billiard boundary, so one applies Dirichlet boundary 
conditions to the billiard walls of a priori unknown shape.  (Such situations 
are known as \begin{itshape}free-boundary problems\end{itshape}.\cite{free})  
The billiard resides in a potential $V$, which adds a second component to the 
electronic Hamiltonian.  Harmonic potentials have been considered most often, 
although quartic ones have been studied a bit as well.\cite{bif}  One observes 
bifurcations in the dynamics as one alters the potential.\cite{bif}  In these 
studies, the potential $V$ depends on the vibrational coordinates $Q$, which 
represent the boundary components undergoing oscillations.  (For the radially 
vibrating spherical quantum billiard, for example, the oscillating portion of 
the boundary is simply the radius $a$.)  Moreover, when $V$ depends implictly 
on time (via the nuclear coordinates), it is considered part of the boundary 
conditions of the problem.  In this situation, the Hamiltonian inserted into 
the Schr\"odinger equation is just the kinetic energy $K$.  If, however, the 
potential $V$ depends explicitly on the spatial variables $(x,y,z)$ or on 
time, the Hamiltonian inserted into Schr\"odinger's equation is instead $K + 
V$.  Finally, note that the electronic variables in vibrating quantum 
billiards may be parametrized by either action-angle variables or Bloch 
variables.\cite{bloch,sazim,ezra}

	In general, one may derive coupled vibrational equations in two 
different manners.  One way is to use the so-called 
\begin{itshape}diabatic\end{itshape} basis.  The molecular wavefunction is 
given by
\begin{equation}
	\psi_d(q,Q) = \sum_k \varphi_k(q;Q_0)\chi_k(Q),
\end{equation}
where the orthonormal electronic states $\varphi_k$ are calculated by solving 
the electronic Schr\"odinger equation at a chosen reference configuration 
$Q_0$:
\begin{equation}
	H_e(Q_0)\varphi_k(q) = E_k^0\varphi_k(q).
\end{equation}

	One then determines the vibrational wavefunctions $\chi_k$ by a set of 
coupled equations with Hamiltonian matrix elements given by
\begin{equation}
	H_{kk'} = T_N\delta_{kk'} + 
\langle\varphi_k|H_e(Q)|\varphi_{k'}\rangle.
\end{equation}
The nuclear kinetic energy $T_N$ is diagonal in this basis, yielding a 
condition that must be satisfied by any physical problem that the present 
analysis is purported to model.  One continues to expand the matrix elements 
of the electronic Hamiltonian $H_e(Q)$ to obtain
\begin{equation}
	H_{kk'} = (T_N + E_k^0 + \Delta U_{NN})\delta_{kk'} + 
\langle\varphi_k|\Delta U_{eN}|\varphi_{k'}\rangle,	
\end{equation}
so that each diagonal element defines an effective vibrational Hamiltonian.  
This Hamiltonian is the sum of the nuclear kinetic energy operator $T_N$ and 
the Hellman-Feynman potential for nuclear motion,\cite{ezra,feynpot} which 
contains a term from internuclear repulsions $U_{NN}$ as well as one from 
attraction to the electronic charge distribution $\varphi_k*\varphi_k$.  Such 
a vibrational Hamiltonian neglects any response that the electronic state may 
have to the changing nuclear configuration.  The off-diagonal coupling terms 
that have been neglected arise from the $\Delta U_{eN}$ term.  Hence, it is 
the change in the potential describing electronic-nuclear attraction as a 
function of the changing nuclear configuration that induces mixing in diabatic
 basis states.  This yields both adiabatic and nonadiabatic correlations of 
electronic and nuclear motion.\cite{ezra}

	Alternatively, one may expand $\psi_d(q,Q)$ using a basis of adiabatic
 electronic states:
\begin{equation}
	\psi_d(q,Q) = \sum_m \varphi_m(q;Q)\chi_m(Q), \label{huang}
\end{equation}
where $\varphi_m(q;Q)$ is a solution of the electronic Schr\"odinger equation
\begin{equation}
	H_e(Q)\psi_m(q;Q) = E_m(Q)\psi_m(q;Q),
\end{equation}
which depends on the nuclear coordinates $Q$.  This is the approach that has 
been followed in the study of vibrating quantum billiards.\cite{sazim,nec,ksu}
  Equation (\ref{huang}) is the Born-Huang expansion, which consists of the 
Born-Oppenheimer expansion plus the diagonal nuclear nonadiabatic coupling.  
The electronic eigenvalues (which are different for different nuclear 
configurations) determine the adiabatic potential surfaces, which change with 
the nuclear configuration because of the dependence of the eigenenergies on 
the nuclear variables $Q$.  One derives equations of motion for the adiabatic 
vibrational amplitudes $\chi_m$ in which the coupling is now due to the 
off-diagonal matrix elements of the nuclear kinetic energy.  

	Translating to mathematical language, we derive a $d$-mode Gal\"erkin 
projection,\cite{infinite} where the integer $d$ refers to the assumption of a
 $d$-level electronic system (corresponding to a $d$-fold near-degeneracy).  
For vibrating quantum billiards, this Born-Oppenheimer expansion corresponds 
to performing an eigenfunction expansion in the wavefunction as though one had
 a stationary boundary (such as a sphere of constant radius if one has 
spherical symmetry) and then reinserting the time-dependence in the resulting 
eigenstates and normalization factors.  For example, this expansion would be 
performed using spherical Bessel functions if one were considering the 
radially vibrating sphere.\cite{sazim}  We note that diabatic expansions are 
often more convenient for practical calculations because they correspond to 
fixed electronic states.  Adiabatic expansions (and associated potential 
energy surfaces), on the other hand, arise naturally from quantum chemistry 
calculations and are also amenable to a dynamical systems approach.

	The analysis of molecular bound states is a particular example 
relevant to the above discussion.  In this situation, the adiabatic potential 
has minima corresponding to nuclear equilibria.\cite{ezra}  (This set of 
minima is not unique, because the system is invariant under spatial 
translations of the molecule as well as rotations about its center of mass.)  
To describe nuclear motion, one has to separate the vibrational 
degrees-of-freedoms from the translational and rotational ones.  (This can be 
formalized mathematically as a \begin{itshape}reduction\end{itshape} 
procedure.\cite{ms})  The potential energy has minima in the vibrational 
coordinates so obtained.  One expands the potential about the equilibrium 
nuclear separations in order to calculate the energies associated with nuclear
 vibration.  The vibrational part of the Schr\"odinger equation ultimately 
becomes a set of coupled harmonic oscillators, which one then studies using 
normal mode expansions.   

	Finally, note that geometric phase (``Berry phase'') often occurs in 
the context of the Born-Oppenheimer approximation.  For the present 
discussion, we (mostly) follow the presentation of Zwanziger, Koenig, and 
Pines.\cite{berry}  Consider a Hamiltonian with two or more variables that 
depend slowly on time.  Thus, we require a 2+ \begin{itshape}dof\end{itshape} 
Hamiltonian system in which at least one of the degrees-of-freedom is slow.  
Semiquantal problems such as those described in the present paper fit into 
this framework.  At each instant, define a smoothly varying (and 
single-valued) basis of eigenstates $\left\{|\psi_k[x(t)]\rangle\right\}$ as 
solutions to the eigenvalue equation
\begin{equation}
	H[x(t)]\left|\psi_k[x(t)]\right\rangle = \lambda_k[x(t)] 
\left|\psi_k[x(t)]\right\rangle.
\end{equation}

	With the adiabatic approximation, a system beginning in the state 
$\left |\psi_k\left[x(0)\right]\right\rangle$ evolves to the state 
$\left|\psi_k\left[x(t)\right]\right\rangle \equiv 
\left|\psi_k\left[x(t)\right]\right\rangle$, which specifies the state at time
 $t$ up to a phase.  However, one must still compute the phase at this time 
relative to that at time zero.  In determining this phase, one must satisfy 
the relation
\begin{equation}
	\left\langle \psi_n\left|\frac{d \psi_n}{d t}\right.\right\rangle = 0,
 \label{ad}
\end{equation}
which can be met at any specific time but which is not necessarily satisfied 
simultaneously at every point in space.\cite{berry}  (In the language of 
geometry, equation (\ref{ad}) gives the requirement for 
\begin{itshape}parallel transport\end{itshape} of the 
\begin{itshape}connection\end{itshape} A, which is a gauge 
shift.\cite{berry,ms,mta,riemann})  That is, this choice in phase may be 
different at different points in space.    Because of this complication in 
defining the phase of the basis 
$\left\{\left|\psi\left[x(t)\right]\right\rangle\right\}$ globally, there 
necessarily exists a phase factor due to the geometry of configuration space 
rather than simply the dynamical equations.  In order to study such 
\begin{itshape}geometric phases\end{itshape}, Sir Michael Berry considered a 
cyclic evolution of period $T$ in the position variables.  The initial and 
final eigenspaces are hence the same, and the problem of comparing the phases 
at times $0$ and $T$ is well-posed. 

	The eigenstate at time $T$ is given by
\begin{equation}
	\left|\psi_n\left[x(T)\right]\right\rangle = 
e^{-i\int_0^T \lambda_n\left[x(t)\right] dt} e^{i\gamma_n(C)} 
\left|\psi_n\left[x(0)\right]\right\rangle,
\end{equation}
where $C$ denotes the closed path traversed in configuration space.  The 
dynamical phase is 
\begin{equation}
	\varsigma_n = \int_0^T \lambda_n\left[x(t)\right]dt,
\end{equation}
and the geometric phase is $\gamma_n(C)$.  Using the Schr\"odinger equation, 
we find that Berry's phase is given by
\begin{equation}
	\gamma_n(C) = \int_C A \cdot dx, \label{ph}
\end{equation}
where
\begin{equation}
	A = \left\langle \psi_n | i \nabla_x \psi_n\right\rangle 
\label{element}
\end{equation}
is the gauge shift.

	The geometric phase $\gamma_n(C)$, which is a real quantity, depends 
only on the initial eigenstate, the geometry of the path $C$, and whether or 
not $C$ surrounds a singularity.  If the configuration space is simply 
connected and $\left|\psi_n[x(t)]\right\rangle$ may be smoothly chosen to be 
real-valued everywhere, then the geometric phase is zero.  We remark that 
equation (\ref{ph}) involves the matrix element (\ref{element}) that causes 
the adiabatic theorem (\ref{ad}) to break down.\cite{berry}  One sees this 
immediately by applying the Chain Rule:
\begin{equation}
	\left\langle \psi_m \left |\frac{d \psi_m}{dt} \right.\right\rangle = 
\left\langle \psi_m \left | \nabla_x \psi_m \cdot \frac{dx}{dt}\right.\right\rangle = 
\left\langle \psi_m \left | \nabla_x \psi_m \right.\right\rangle \cdot \frac{dx}{dt}.
\end{equation}
Also note that one may derive Berry's phase using a path integral formulation.
  To consider the relevance of geometric phase to adiabatically evolving 
systems, we remove the $\exp[i \gamma_n(C)]$ factor via a gauge 
transformation.  This adjusts our Hamiltonian by adding a vector potential 
term given by equation (\ref{element}).  A system with nonvanishing Berry 
phase may thus be treated using an effective Hamiltonian obtained via the 
transformation
\begin{equation}
	\nabla \mapsto \nabla - iA. \label{eff}
\end{equation}
With equation (\ref{eff}), the geometric phase is absorbed into the dynamical 
phase $\varsigma_n$, even though its ultimate source is the geometry of 
configuration space.  

	The gauge potential just discussed is Abelian, like the one that
 occurs in electromagnetism.\cite{qft}  Although it occurs in an abstract 
space, one may still observe its effects.  As in more familiar situations, it 
arises from ambiguities in the description of a system.  In the present case, 
these ambiguities reflect the variety of manners in which one may select 
eigenstate phases.  Such an ambiguity in assigning relative coordinates occurs
 whenever a problem may be separated into two subcomponents.  In the present 
context, one has a natural separation into fast (electronic) and slow 
(nuclear) dynamics.  In principle, one may therefore derive a geometric phase 
for any vibrating quantum billiard.  Such analyses have already been performed
 for similar molecular models such as the $E \times e$ Jahn-Teller system.  
One computes a geometric phase $\gamma(C)$ of value $-\pi$, yielding a phase 
factor of -1.  Hence, if one specifies a basis of single-valued eigenstates, 
Berry's phase imposes a sign change and double-valued behavior.  One may treat
 this example as a special case of a three-dimensional problem such as a spin 
1/2 particle in a planar magnetic field.  (That is, the particle may move in 
three-space, but the field is planar.)  The mathematical distinction between 
this perspective and the previous one is that analyzing the problem in 
two-dimensions forces a topological interpretation of the sign change--whereas
 considering a third dimension permits a geometric interpretation of the sign 
change.  Interpreted topologically, the sign change arises from the fact that 
the configuration space $(\mathbb{R}^2 -\{0\})$ is not simply connected.  
Interpreted geometrically, the configuration space becomes curved because it 
deviates from the plane.  In exchange, it is now simply connected.  This 
geometric perspective illustrates a similarity between molecular spectroscopy 
and spin in a magnetic field.\cite{berry}  Both the Jahn-Teller $E \times e$ 
problem and the spin 1/2 particle in a planar magnetic field are described by 
real, symmetric $2 \times 2$ matrices, so the fundamental similarities between
 these systems are ultimately due to the commonlity of their Lie 
structure.\cite{ms}  One may think of the magnetic field as representing the 
semiclassical limit of a localized packet of nuclear configurations, and the 
corresponding molecular problem may be generalized to three dimensions by 
analyzing molecules whose nuclear configurations have higher symmetries.  One 
may thus construct an analogy between vibrations in molecules and crystals and
 the problem of a spin 1/2 particle in a magnetic field.  Additionally, one 
may construct analogies with other spin systems.  For example, the triply 
degenerate electronic state of a $T \times (t + e)$ system is related to a 
spin 1 particle in combined magnetic and electric quadrupole fields.  This 
idea of analyzing systems based on their common symmetries is a hallmark of 
the field of geometric mechanics,\cite{ms} which may prove very useful to the 
study of molecular dynamics as well.  

\vspace{.05 in}

\begin{centering}
\section{Molecules and Crystals}
\end{centering}

	In the present section, we discuss why polyatomic molecules are 
appropriately described semiquantally.  We hinted at this description earlier,
 but we now consider it at length.  Before beginning this discussion, however,
 it is important to recall that--in addition to the modern results reviewed in
 the present paper--molecules and crystals can also be studied using 
well-known techniques such as the Bloch theorem.\cite{liboff,sakurai,group}

	Earlier in this work, we noted that polyatomic molecules may be 
abstracted as vibrating quantum billiards.  This abstraction arose initially 
by considering one-dimensional vibrating quantum billiards as a model for 
diatomic molecules.\cite{vibline}  More generally, a $d$-mode Gal\"erkin 
expansion of a vibrating quantum billiard corresponds to a $d$-term quantum 
system coupled in a time-dependent, self-consistent fashion to $r$ classical 
degrees-of-freedom.  These classical \begin{itshape}dof\end{itshape} 
correspond to the \begin{itshape}dov\end{itshape} of the quantum billiard. 
 Taking $d = 2$ and $r = 1$, one obtains a system precisely analogous to a 
diatomic molecule (such as NaCl) with two electronic states (of the same 
symmetry) coupled nonadiabatically by a single internuclear vibrational 
coordinate.  Such symmetry, which can be expressed in terms of
 symmetry conditions in a superposition state's quantum 
numbers,\cite{sazim,nec} are one possible cause of the electronic degeneracy 
discussed earlier.  That is, degeneracy and near-degeneracy of electronic 
energy levels are often a consequence of a system's inherent symmetries.  
(It is important to note that there are also many systems in which the 
near-degeneracy of electronic eigenenergies is \begin{itshape}not\end{itshape}
 caused by symmetry.)

	Vibrations in more complicated molecules can have more nuclear 
degrees-of-freedom, although because of constraints, they might not all be 
independent.  Thus, only polyatomic molecules of certain forms are describable
 directly as vibrating quantum billiards.  Others may be described in terms of
 other semiquantal models that depart somewhat from the vibrating billiard 
abstraction.  These models nevertheless retain a semiquantal formulation 
analogous to that of vibrating quantum billiards.  Additionally, there exist 
molecular systems for which the vibrating quantum billiard model is apt 
without adjustment.  A two \begin{itshape}dov\end{itshape} quantum billiard 
such as one with rectangular geometry with length $a$ and width $b$, for 
example, may be used to model nonadiabatic dynamics in a linear triatomic 
molecule with respective nuclear displacements $a(t)$ and $b(t)$ between the 
center atom and the two atoms to which it is bonded.  Requiring left and right
 symmetry for all time--represented by the constraint $a(t) \equiv 
b(t)$--yields the quantum vibrating square, which has one 
\begin{itshape}dov.\end{itshape}\cite{rect}  An $r$ 
\begin{itshape}dov\end{itshape} quantum billiard may likewise be used to model
 nonadiabatic coupling between the nuclear and electronic states of a linear 
molecule consisting of a chain of $(r + 1)$ atoms.  Perhaps more physically 
relevant would be using an $r$ \begin{itshape}dov\end{itshape} quantum 
billiard as a model for the coupling between electronic states and $r$ normal 
modes in a polyatomic molecule such as benzene.  One can complicate the 
situation further by considering the coupling of polyatomic molecules by 
intermolecular forces such as hydrogen bonds.  Such systems are describable as 
coupled vibrating quantum billiards and hence as coupled Schr\"odinger 
equations with time-dependent boundary conditions.  To our knowledge, these 
systems have yet to be studied in this fashion.

	For electronic-nuclear coupling, it is simplest to consider two-fold 
degeneracies, for which one uses two-mode Gal\"erkin expansions.  An example 
of such a degeneracy occurs in the Jahn-Teller $E \times e$ trigonal molecule,
 in which one doubly degenerate vibrational mode ($e$) interacts with a 
(symmetry-induced) double electronic degeneracy ($E$).  This system has been 
analyzed by Whetten, Ezra, and Grant\cite{ezra} and by Zwanziger, Grant, and 
Ezra\cite{ezra2}.  Let us back up a bit, however, and give a physical 
discussion of why Hamiltonian models of electronic-nuclear coupling are good 
qualitative approximations of reality.  Note again that there are some systems
 for which even the abstract vibrating quantum billiard formulation (which is 
a toy model) is useful in this fashion.

	The easiest situations to understand are ones in which the degenerate 
electronic energy levels are isolated energetically from other states.  That 
way, one may neglect these other states, providing justification for the use 
of low-mode Gal\"erkin expansions.  Additionally, we assume that 
the molecule under consideration possesses strong restoring forces that 
prevent large-amplitude nuclear motions.  Systems with these properties 
ordinarily arise from single-hole or single-electron degeneracies.  One then 
examines the system quantum-mechanically in order to examine the coupling of 
electronic, vibrational, and rotational motion in these systems.  One can 
examine the distribution of energy levels, for example, by considering 
correlation functions and limiting situations.

	A hallmark of electronic degeneracy is the extreme sensitivity of 
rovibrational states to perturbations.  The nuclear motions of a molecule and 
its coupling to electronic motion may be influenced heavily by the surrounding
 environment.  Hence, in order to study individual rovibrational states 
(rather than the bulk properties of nonadiabatic systems of bound states), one
 must isolate the molecule of interest.  This is a highly nontrivial 
proposition.  Moreover, strong nonadiabaticity is rare in the lower 
rovibrational levels of ground-state electronic terms of most easily isolated 
molecules.\cite{ezra}  For example, the ground states of stable molecules 
contain an even number of electrons with electronic components that are 
ordinarily fully symmetric.  Open-shell polyatomic molecules with sufficient 
symmetry often also exhibit electronic degeneracies, but it is difficult to 
prepare such systems.  Many transition metals have degenerate ground terms, 
although they are usually only observed in solutions or solids.  At higher 
energies, such as those near the threshholds for chemical reactions or 
molecular dissociation, one expects to observe strong rovibrational motion.  
These situations, however, are not highly symmetric.  They are also infested 
by complex, large-amplitude nuclear oscillations.  Consequently, one cannot 
expect to examine the ground terms of stable molecules in order to 
experimentally observe few-state nonadiabatic motion.

	Another situation worth considering involves excited molecular valence
 states, which display strong interactions between two or more electrons or 
holes.  Many of the symmetric situations contain degenerate (or nearly 
degenerate) electronic energy levels.  Spectroscopic examination of such 
systems often reveals degenerate molecular orbitals, so the excited states 
contain four or more related terms (of which two are degenerate).  One can 
thus describe this electronic-nuclear coupling with a superposition state of 
four or more terms.  One can similarly consider the ground states of molecules
 such as cyclobutadiene and cyclo-octatetraene, in which the two electrons of 
highest excitation arise from the the same degenerate molecular 
orbital.\cite{ezra}

	A related application is the analysis of Jahn-Teller systems.  In the 
presentation below, we borrow material from \begin{itshape}Vibronic 
Interactions in Molecules and Crystals\end{itshape} by I. B. Bersuker and V. 
Z. Polinger.\cite{vibe}  The \begin{itshape}Jahn-Teller theorem\end{itshape}, 
which may be treated mathematically using the language of algebra and 
functional analysis, states that if the adiabatic potential of a system (which
 is a formal solution to the electronic part of the Schr\"odinger equation) 
has several crossing sheets, then at least one of these sheets has no extremum
 at the crossing point.  Hence, degenerate (and even near-degenerate) 
electronic energy levels cannot be analyzed using the Born-Oppenheimer 
(adiabatic) approximation.  As this approximation breaks down, one obtains 
nonadiabatic coupling between nuclear and electronic terms.  That is, near 
such degeneracies and near-degeneracies, it is appropriate to use the 
semiquantal regime in order to study the dependence of the electronic 
eigenenergies on the system's nuclear degrees-of-freedom.  According to the 
semiquantal approximation, this dependence is approximated as a continuum.  

	It may be apt to use the Jahn-Teller effect as a 
\begin{itshape}synonym\end{itshape} for nonadiabatic coupling between nuclear 
and electronic systems, but it is conventional to apply this term more 
specifically.  Hence, although vibrating quantum billiards exhibit behavior 
like the Jahn-Teller effect, it is not directly labeled as such.  
Additionally, we note that in the case of electronic near-degeneracies, the 
term \begin{itshape}pseudo-Jahn-Teller effect\end{itshape} is sometimes used. 
 Situations in which Jahn-Teller deformations have been observed include 
vibrations in crystals, numerous types of spectroscopy (NMR, Raman, 
\begin{itshape}etc.\end{itshape}), multipole moments, the stereochemistry and 
instability of molecules, mechanisms of chemical reactions, and catalysis.  
Crystals that exhibit the Jahn-Teller effect (such as ferroelectric crystals) 
contrast strikingly with those that do not.  For example, their structural 
phase transitions and elastic properties are different.\cite{vibe}

	Effects analogous to Jahn-Teller distortion have been observed in 
other physical systems, including the pion-nucleon interaction in quantum 
field theory, the $\alpha$-cluster description of light nuclei, and the 
resonant interaction of light with matter.  The formal analogy between 
pion-nucleon interaction in the static model of the nucleon\cite{corn2} and 
the Jahn-Teller problem is used to apply the methods and ideas of scattering 
theory to Jahn-Teller distortions.\cite{vibe}  One can also apply this analogy 
in reverse to study semiquantum chaos and nonadiabatic dynamics in 
pion-nucleon interactions, a special form of Yukawa coupling in quantum 
chromodynamics (QCD).\cite{qft}  To our knowledge, this has not yet been done.
  Indeed, any system that exhibits Jahn-Teller deformation is expected to 
exhibit semiquantum chaos.  If the system has three or more 
degrees-of-freedom, it may also exhibit semiquantum 
diffusion.\cite{lich,haller}  

	Jahn-Teller systems exhibit equivalent minima of adiabatic surfaces 
which correspond to several distorted nuclear configurations of equivalent 
symmetry.  For example, a molecule of type $ML_6$ with a double electronic 
degeneracy in the regular octahedral configuration becomes elongated along one
 of its four-fold axes of symmetry $C_4$ because of the Jahn-Teller effect.  
There are three equivalent distortions because there are three axes with $C_4$
 symmetry.  (The set $C_m$ denotes the group of rotational symmetries of the 
$m$-gon.\cite{group,dummit}  Its elements are rotations by angle $2\pi/m$, 
which are denoted $c_m$.)  The Jahn-Teller theorem does not apply to double 
electronic spin degeneracies (\begin{itshape}Kramers 
degeneracies\end{itshape}) or linear molecules.  However, linear molecules 
with degenerate electronic eigenenergies are unstable with respect to bending 
distortions, which produces nonadiabatic behavior known as the 
\begin{itshape}Renner effect.\end{itshape}  Additionally, one 
\begin{itshape}dov\end{itshape} quantum billiards may be used to describe 
nonadiabatic dynamics in diatomic (and hence linear) molecules.  Though 
related to the Jahn-Teller effect, this nonadiabatic behavior is not precisely
 the same.  It is more accurately termed the \begin{itshape}inverse 
Jahn-Teller effect\end{itshape} because of the inverted dependence on the 
interatomic distance $a$.  This leads to degeneracy in the adiabatic sheets at
 $a = \infty$ rather than at $a = 0$ as in the Jahn-Teller effect.  The 
consequences of this degeneracy may be observed in diatomic molecules when 
their interatomic displacement is large.  As we have been stressing, 
electronic near-degeneracies are as important as actual degeneracies.  In 
these near-degeneracies, two potential surfaces almost intersect but do not 
actually cross each other because of some sort of weak interaction near at the 
point of closest approach.  The observation that such near-degeneracies are 
prevalent in diatomic molecules has been captured by the one 
\begin{itshape}dov\end{itshape} quantum billiard model.  An example of this 
phenomenon is in the near-intersection between the lowest ionic and covalent 
states of alkali halides.  In NaCl, this near-degeneracy occurs at large 
interatomic distances--which is consistent with our predictions.  When the 
sodium and chloride ions are far apart from each other, the resonance integral
 for electron transfer between them is small.  Hence, the smallness of the 
cross-term $H_{12}$ in the Hamiltonian is caused by the large difference in 
the electron distribution in the two electronic states.  Analogous situations 
occur in more complicated polyatomic molecules.\cite{intersect}

	Jahn-Teller effects are classified according to their tensorial 
construction, which describes their (nuclear and electronic) symmetries and 
degeneracies.  The $E$ term refers to an orbital doublet (electronic double 
degeneracy).  The canonical $E \times e$ Jahn-Teller effect describes the 
interaction of doubly-degenerate electronic states of representation $E_k$ 
term with doubly-degenerate nuclear vibrations of representation 
$E_{2k}$.\cite{vibe}  (The representation $E_m$ has basis functions with the 
transformation properties $\psi_{\pm m} \sim \exp(\pm m \varphi)$, where 
$\varphi$ is an arbitrary rotation about the symmetry axis.)  The simplest 
polyatomic systems in which this occurs are triangular molecules $X_3$, 
tetrahedral molecules $ML_4$, and octahedral molecules $ML_6$.  The presence 
of nuclear degeneracies of this type lead to multmode $E \times (e + \cdots 
e)$ systems, where the number of factors of $e$ correspond to the number of 
vibrational modes of type $E$.  

	Another orbital-doublet Jahn-Teller system is denoted $E \times (b_1 +
 b_2)$, which occur when electronic states transform as $\psi_{\pm m} \sim 
\exp(\pm i m \varphi/4)$ in polyatomic systems with $C_m$ or $S_m$ ($m = 4k$) 
axes of symmetry.  (The set $S_m$ denotes the group of \begin{itshape}rotary 
reflections\end{itshape} in the $m$-gon.  Its elements are given by $s_m 
\equiv \sigma_h c_m = p_0 c_2 c_m$, where $p_0$ represents parity (spatial 
inversion) and $\sigma_h$ is mirror reflection in the horizontal 
plane.\cite{group}  Rotary reflections are examples of \begin{itshape}improper
 rotations.\end{itshape})  In the present situation, the active Jahn-Teller 
modes are the singlet low-symmetry displacements (one-dimensional 
representations) of type $B_1$ and $B_2$.  Each $B_i$ represents an 
independent, nondegenerate vibrational mode.  The simplest examples of $E 
\times (b_1 + b_2)$ distortions occur in square planar and pyramidal molecules
 $ML_4$.  One special case of interest, denoted $E \times b$, occurs when the 
vibronic coupling to one of the $B$ vibrational modes is neglible.  The motion
 along the coordinate decoupled from the electronic 
\begin{itshape}dof\end{itshape} consists of simple harmonic vibrations near 
the completely symmetric nuclear configuration.  

	The $T \times (e + t_2)$ Jahn-Teller distortion refers to the coupling
 of a triple electronic degeneracy $T$ with a doubly-degenerate nuclear 
vibration of type $E$ and a triply-degenerate nuclear mode of type $T$.  The 
simplest cubic molecule that can exhibit such a distortion is $ML_6$, whose 
symmetry group is $O_h$.  (The octahedral group $O_h = O \times C_i$ describes
 the full symmetry of a cube.  The group $O$, consisting of 24 rotations, is 
the set of rotational symmetries of cubes and regular octahedrons.  The group 
$C_i$ consists of the parity operator $p_0$ and the identity $I$.)  Some 
important special cases are the $T \times e$, $T \times t_2$, $T \times d$, 
and $P \times d$ Jahn-Teller distortions.  The adiabatic potential in the 
$T_2$ subspace has $O \cong T_d$ symmetry, where $T_d$ represents the full 
symmetry group of a tetrahedron.  Type $D$ nuclear distortions describe the 
case of equal coupling of the electronic \begin{itshape}dof\end{itshape} to 
the $E$ and $T_2$ nuclear modes at equal forcing $\omega \equiv \omega_E = 
\omega_T$ when the molecular Hamiltonian (describing the coupling between 
electronic and vibrational degrees-of-freedom) posesses $S0(3)$ symmetry.  
(The parameter $\omega^2$ denotes the force constant of the normal vibration.)
  The type $P$ triple electronic degeneracy exhibits more symmetries than the 
type $T$ degeneracy.  Therefore, the $P \times d$ distortion is a highly 
symmetric special case of the $T \times (e + t_2)$ Jahn-Teller systems.  It 
occurs in cubic polyatomic systems in which the cubic splitting of $D$ mode of
 nuclear vibrations is negligible.\cite{vibe}  That is, the $P$ configuration 
refers to three-fold degenerate $p$-orbitals $(j = 0, m \in \{0, \pm 1\})$, 
and the $D$ configuration represents five-fold degenerate $d$-orbitals $(j = 
2, m \in \{0, \pm 1, \pm 2\})$.  (Recall that $l$ and $m$ denote the orbital 
and azimuthal quantum numbers, respectively.\cite{liboff})  The $P \times d$ 
Jahn-Teller distortion then represents a triply degenerate nuclear vibration 
coupled to a five-dimensional electronic manifold in spherical symmetry.  In 
going from spherical symmetry to octahedral symmetry, the $d$ orbitals split 
into $(e + t_2)$.  The same coupling to $e$ and $t$ modes would then indicate 
a remnant of higher (spherical) symmetry.\cite{group}

	Other Jahn-Teller effects include the $\Gamma_8 \times (e + t_2)$ 
distortion and its special cases.  Quadruplet terms (such as $\Gamma_8$) occur
 in icosohedral systems as well as in cubic systems with spin-orbital 
coupling.  (The term $\Gamma_8$ is the double-valued irreducible 
representation  of the double group $\bar{O}^2$ of the rotation group of 
regular octahedrons $O$.  \begin{itshape}Double groups\end{itshape} are 
obtained by treating rotations by the angles $\alpha$ and $(\alpha + 2\pi)$ as
 different quantities even though they describe physically identical 
rotations.\cite{group}) The simplest situations with electronic quadruplets 
(corresponding to four-mode Gal\"erkin expansions) occur in cubic polyatomic 
molecules which have an odd number of electrons.  Icosohedral systems can also
 exhibit vibronic coupling of nuclear quadruplets and quintuplets (of types 
$u$ and $v$, respectively) with electronic quintuplets and quintuplets (of 
types $U$ and $V$, respectively).  One possible interaction of these terms is 
$U \times (u + v)$, in which a quadruply degenerate electronic state is 
coupled both to quadruply and quintuply degenerate nuclear modes.\cite{vibe}  

	One can consider arbitrarily complicated Jahn-Teller distortions, even
 the simplest of which are dynamically interesting.  Multiple distortions can 
occur in the same molecule or crystal.  The individual distortions need not 
possess the same symmetry.  Systems that exhibit multiple sets of Jahn-Teller 
distortions of a given symmetry are called \begin{itshape}multimode 
Jahn-Teller systems\end{itshape}.  Multiple distortions can occur, for 
example, in crystals with point defects, as their energy spectra contain 
discrete near-degenerate electronic eigenenergies well-separated from other 
energy levels.  One can also observe multiple Jahn-Teller centers interacting 
with each other.  Systems that exhibit such behavior are known as 
\begin{itshape}polynuclear clusters\end{itshape}.  For example, consider two 
octahedral complexes of type $ML_6$ can form a double-center Jahn-Teller 
system (a bioctahedron) in three different ways with the two central atoms 
lying on the common axes of symmetry of the second, third, or fourth order.

	Both theoretical and experimental analyses have been vital to the 
study of nuclear-electronic coupling in molecules.  Treatments with few active
 modes--that is, low-mode Gal\"erkin projections--have been particularly 
useful, because they ease the analytical difficulty of the theory.  One may 
then consider a fully quantum variational treatment of the nuclear part of the
 Hamiltonian, so that model parameters can be fit to experimental data.  (One 
molecule for which this has been done is 
\begin{itshape}sym\end{itshape}-triazine.\cite{ezra})  As one increases the 
order of the Gal\"erkin approximation, however, the procedure becomes 
increasingly difficult both analytically and computationally.  Hence, treating
 these situations in the fully quantum regime becomes untenable rather
 quickly.  It is consequently useful to develop semiclassical and semiquantal 
techniques even for few-term superposition states in order to analyze 
molecular systems.  

	The present work describes a semiquantal technique, but a 
semiclassical theory (using the $\hbar \longrightarrow 0$ asymptotic 
formalism) of several-mode nuclear motion is worth studying as well.  (Recall 
that the semiquantal regime is a semiclassical regime but not 
\begin{itshape}the\end{itshape} ``semiclassical'' regime that is traditionally
 studied as part of the quantum mechanics curriculum.)  If one considers this 
situation abstractly in terms of vibrating quantum billiards, then a 
semiclassical procedure does exist in principle.  One quantizes the motion of 
the billiard boundary to obtain an example of quantum chaology (for which one 
uses the semiclassical limiting procedure).  To our knowledge, however, this 
process has not actually been carried out for billiards with oscillating 
boundaries or for the physical situations described in the present paper.

\begin{centering}
\section{Solvated Electrons}
\end{centering}

\vspace{.05 in}

	Vibrating quantum billiards may also be useful for the study of 
electron solvation, which describes the nonadiabatic process of the relaxation
 of excess electrons in fluids.  In this discussion, we adapt the work of 
Space and Coker to this perspective.\cite{solve1,solve2}  The equilibrium 
structure of electron solvation in numerous fluids involves the localization 
of excess electrons into a roughly spherical cavity within the solvent.  This 
cavity is the boundary of our billiard, so describing the present situation 
with the radially vibrating spherical quantum billiard may be especially apt. 
 The electron is treated quantum-mechanically, whereas the motion of the 
solvent is approximated as classical.  Such systems are hence in the confines 
of the semiquantal regime.  Short-range, repulsive electron-solvent 
interactions prevent solvent molecules from penetrating the billiard boundary 
into the region occupied by the electron.  

	There are a wide variety of relevant fluid types.  Among the 
possibilities are simple hard-sphere liquids such as helium and exceptionally 
polar solvents like water and ammonia.  Calculations have explored the 
dynamical rearrangements that solvent atoms must make in order to accomodate 
an injected electron as well as how the electron and solvent couple to provide
 various relaxation pathways.  This description is consistent with recent 
analyses of vibrating quantum billiards.\cite{sazim,ksu,nec,rect}

	Equlibrium excess electronic states possess fairly well-defined 
symmetry\cite{solve2}, but solvated electronic states in an unperturbed fluid 
are more disordered and delocalized, as there is no well sufficiently deep to 
serve as an equilibrium trap.  The electron has numerous near-degenerate local
 minima, so the system's eigenstates are combinations of functions centered in
 several different wells.  States may thus have a positive density in more 
than one cavity, although the ground electronic state is usually in an 
$s$-like state of a single cavity.  Recall that the canonical vibrating 
billiard model assumes boundary fluctuations that preserve some or all of the 
symmetries of the billiard.  We thus surmise that such a model may be 
appropriate for equilibrium fluctuations, because in that situation one has a 
single cavity with many symmetries.  The simplest model to use in this event 
would be the radially vibrating spherical quantum billiard.  In order to 
account for the observed asymmetry prior to equilibrium, one may be able to 
extend the model by considering multiple-well potentials, higher-term 
Gal\"erkin projections (which would account for larger degeneracies), multiple
 degrees-of-vibration, and--perhaps most importantly--coupled vibrating 
quantum billiards.  All except the latter may be inserted into the model for 
a single vibrating billiard to obtain a more complicated single-cavity 
configuration.  One would need to consider coupled billiards in order to 
capture the notion of multiple cavities. 

	When a solvated electron in the ground state of a solvent cavity is 
excited by light (``photoexcited'') into various electronic states, one 
obtains nonadiabatic relaxation, which is suitably described using the 
semiquantal approximation.  One may also explore the nonadiabatic relaxation 
that occurs when an excess electron is injected into various excited states of
 unperturbed fluid configurations.  With such analysis, one may study the 
dynamical trapping processes that are responsible for the formation of the 
localized equilibrium state (in other words, the quantum billiard system).

	As with polyatomic molecules, one uses the Born-Oppenheimer 
approximation to produce a semiquantal analysis.  One obtains trajectories by 
solving classical equations of motion for the cavity.  These evolution 
equations include forces due to the charge distribution of the currently 
occupied electronic state.  Space and Coker\cite{solve1} have shown that at 
least two types of nonadiabatic processes are important for solvated electron 
relaxation in simple fluids.  The first of these processes is highly diabatic 
and may be illustrated by a pair of weakly coupled $p$-like orbital states in 
the same solvent cavity.  Suppose that a $p_y$ state has a higher energy than 
a $p_x$ state because the cavity is not quite spherical.  The cavity will then
 begin to elongate in the $y$-direction because of the influence of the charge 
distribution, which causes the energy gap between the two states to decrease. 
 If the two states are weakly coupled, they mix only for small separations of 
their energy levels.  Hence, appreciable mixing between these states occurs 
only in a small region of nuclear (solvent) configuration space.  
Consequently, the manifestation of this behavior is brief.  In the present 
example, the diabatic transition occurs when the two adiabatic basis states 
(the two different types of $p$-orbitals) rotate into each other rapidly, 
which causes the occupied electron to hop between the two states.  As a 
result, the occupied electron state, which is still causing the solvent cavity
 to expand in the $y$-direction, is now less energetic than the unoccupied 
$p_x$-orbital.  There is never any drastic change in the electronic charge 
distribution.

	Now consider highly diabatic relaxation in a more general 
context.\cite{solve2}  There is typically a nonzero electron density in more 
than one cavity in the fluid.  This density exerts an outward forces on the 
surrounding solvent.  Additionally, distinct states are coupled 
nonadiabatically (because they each have a density in the same trap).  The 
nuclear velocities are initially uncoupled to the electronic forces, so the 
electron rapidly explores several closely-related regions before localizing in
 a single equilibrium density fluctuation.  In a surface-hopping calculation, 
the excess electron hops between eigenstates that change fairly quickly with 
the solvent configuration (as the wells in the fluid are almost degenerate).  
The transitions in question are ordinarily not simple diabatic ones in which 
state identities change.  Instead, they are strongly nonadiabatic, as the 
wavefunctions become a dynamically changing mixture of several states in 
closely-related solvent regions.  This is a consequence of the lack of 
symmetry in the configuration space and phase space of the initial unperturbed
 solvent.  Once the electron is localized to a single cavity, its transitions 
tend to be diabatic.  It pushes out the surrounding solvent to an equilibrium 
state where the solvent-solvent and electron-solvent forces nearly balance.  
At this point, the canonical vibrating quantum billiard model is especially 
appropriate, as one now has the requisite symmetry preservation.

	A second important nonadiabatic process occurs when two interacting 
electronic states are strongly coupled by the solvent dynamics.  Suppose, for 
example, that the $p_y$ solvated-electron charge distribution considered above
 has continued its expansion of the solvent cavity along the $y$-axis.  
Simultaneously, solvent atoms near the cavity walls around the node in the 
$p_y$ wavefunction have begun to push into the cavity, which pinches off the 
middle region of the ground-state wavefunction, thereby causing it to increase
 in energy (since the relevant displacement parameter of the billiard is 
smaller when this happens).  These two electron states occupy similar regions 
of space and are strongly coupled by the motion of the cavity boundary.  The 
two associated energy levels may thus switch their ordering, as they can mix 
strongly even when there is a large energy disparity.  When this occurs, the 
excess electronic potential due to the deforming cavity looks like a 
fluctuating double well.  When the distorted $s$ and $p$ orbitals mix 
strongly, the solvated electron may be localized on either the left or right 
side of the double-well cavity.  Eventually, either a left-well or right-well 
localization predominates, and the electron hops into a ground-state 
$s$-orbital.

	The crux of the matter is that solvated electrons exhibit transient 
relaxation dynamics before they reach their equilibrium behavior.  The 
canonical vibrating quantum billiard model may be useful for the equilibrium 
oscillations, whereas more complicated extensions of the model (especially 
coupled vibrating quantum billiards) may be appropriate to help describe 
several phases of the transient behavior.  At short times, the occupied 
electron state is embedded in a continuum, so if the fluid density is 
sufficiently low, the electron can leak through the solvent and hop from state
 to state.  At this point, the excess electronic absorption is characterized 
by a low-energy band.  Once the solvated electron is in a particular cavity 
for a sufficiently long time interval, the diameter of the cavity increases 
rapidly, thereby localizing the solvated electron in a deeper well.  The wall 
of the cavity still oscillates, but it is now reasonable to assume that its 
geometry is preserved under these vibrations, so that simple toy models such 
as vibrating quantum billiards may be useful.

\begin{centering}
\section{Buckyballs and Nanotubes}
\end{centering}

\vspace{-.5 in}

\begin{centering}
\subsection{Buckyballs}
\end{centering}	

	It is well-known that the motion of electrons through quantum dots may
 be altered substantially by energy-level quantization and the charging of a 
single electron.  Studies of electron transport have recently been extended to
 the realm of chemical nanostructures such as nanocrystals and nanotubes.  A 
group of scientists at Lawrence Berkeley National Laboratory fabricated 
unimolecular transistors by connecting buckyballs to gold 
electrodes.\cite{bucky}  Buckyballs, \begin{itshape}Time 
Magazine's\end{itshape} molecule of the year in 1991, are $C_{60}$ molecules 
shaped like a soccer ball.  The LBNL researchers then studied the 
nanomechanical vibrations of these buckyball transistors.  They performed 
transport measurements that demonstrated coupling between the (nuclear) motion
 of the center-of-mass of the buckyball and the hopping of the single electron
 (electronic motion).  This conduction mechanism was not observed previously 
in quantum dot studies, although such electronic-nuclear coupling is a 
hallmark of nonadiabatic motion in molecular dynamics and the vibrating 
quantum billiard model.

	The $C_{60}$ molecule is a fullerine that--like a soccer ball--has a 
surface consisting of twenty hexagons and twelve pentagons.  Its geometry is 
thus more complicated than those that have been used in the study of vibrating 
quantum billiards, but it exhibits a similar type of nonadiabatic 
electronic-nuclear coupling.  However, one can obtain a simpler model by 
approximating the buckyball as a sphere in order to study the coupling between
 its bouncing and the electronic motion.  Perhaps one could then use 
perturbation theory to account for the more complicated geometry.  Controlling
 such motion (and other types of motion!) in nanoscale objects is an important
 issue in the field of nanotechnology.  Macroscopically, one may ignore the 
coupling of a rigid wall to a ball bouncing against it, but at sufficiently 
small scales one can no longer do this, a fact which is at the heart of the 
present review.   In other words, when one studies objects at ``new'' scales 
(such as microscales and nanoscales), one must examine couplings between types
 of systems that have not traditionally been treated together as subcomponents
 of a single entity.  Electronic-nuclear coupling heavily influences molecular
 motion, even though the electron mass is a small fraction of the molecular 
mass.  The mechanical control of nanoscale objects (``NEMs'') will allow 
smaller, faster, and more efficient versions of existing 
micro-electro-mechanic structures (MEMs).\cite{bounce}  A single-electron 
current can both detect and excite mechanical oscillations in a buckyball.  
For example, an electron with surplus energy precisely equal to the 
vibrational energy of the buckyball causes the buckyball to begin bouncing due
 to spontaneous emission of this energy.  Furthermore, the electron continues 
to hop on and off the molecule.  Electronic devices in which the 
electro-mechanical motion is so coupled could function as ``electron 
turnstiles'' that allow electrons to pass one at a time.

	The various devices studied by Park, \begin{itshape}et al.\end{itshape}
 exhibited a universal quantized excitation with an energy of about 5 meV.  
Such an excitation energy could arise in several possible manners in a 
single $C_{60}$ transistor.  One hypothesis, which has been invoked in other 
nanosystems, is that this excitation energy is a result of highly excited 
electronic states of the buckyball.  However, this possibility was dismissed 
for several reasons.  The observed excitation energy was the same for both 
types of charge states of $C_{60}$, multiple excitations with the same spacing
 were observed, and this mechanism is not consistent with theoretical 
calculations of the electronic states of $C_{60}^{n-}$ ions.  

	A better explanation involves coupling between vibrational excitations
 of the buckyball with electronic tunnelling on and off $C_{60}$.  That is, 
the authors proposed a nonadiabatic semiquantal system in which squishing 
couples to electronic hopping.  In this nuclear mode, the buckyball deforms 
(``squishes'') a little bit like a ball being pressed against the ground.  
With this mechanism, the observation of multiple $\partial I/\partial V$ 
features with identical spacing would then result from the excitation of 
integral numbers of vibrational quanta.  These vibrational modes, moreoever, 
would be identical for both charge states of the buckyball.  However, there 
are some problems with this explanation as well.  The internal vibrational 
mode of lowest energy is about 35 meV, so we require a different explanation. 
 The lowest pulsing internal mode, like what occurs in the radially vibrating 
spherical quantum billiard, is even more energetic. 

	A possible resolution involves another type of electronic-nuclear 
coupling.  One hypothesizes a mechanism in which oscillations of the 
buckyball's center-of-mass within a confinement potential bind it to a gold
 surface, thereby predicting a vibrational energy of about 5eV (as was 
observed experimentally).  Hence, the following nanomechanical motion is 
predicted: an electron jumps onto a $C_{60}^{n-}$ molecule, which causes an 
attraction between the additional electron and its image charge on the gold.  
This electrostatic interaction pulls the buckyball ion closer to the gold 
surface and results in mechanical motion of $C_{60}$, like that of a 
soccerball bouncing against the ground.  Although slightly different from the 
mechanisms we have been discussing (which correspond more closely to the 
second hypothesis), this third hypothesis also predicts nonadiabatic 
electronic-nuclear coupling, for which a semiquantal description is 
appropriate.  It is amenable to the same analyses that have been performed on 
vibrating quantum billiards, but the relevant Hamiltonian takes a slightly 
different form.

	The dynamical situation in the bouncing buckyball is reminiscent of 
the Franck-Condon process that occurs in electron transfer and light 
absorption in molecules.  In these situations, the electronic motion is 
accompanied by vibrational excitation.  The transport measurements discussed 
above may be used both to probe and to excite molecular motion.  In 
particular, the buckyball transistor that was studied behaved as a 
high-frequency nanomechanical oscillator.  The electronic component of its 
oscillations may be quantized so that the system is treatable in the 
semiquantal regime.  This sort of coupling between quantized electronic (fast)
 and unquantized mechanical (slow) degrees-of-freedom should become important 
for electron transport through nanomechanical systems such as buckyballs and 
carbon nanotubes.  The difference between this system and a vibrating quantum 
billiard of appropriate geometry is that the nuclear motion is described by a 
``bouncing'' of the center-of-mass rather than a pulsing of the billiard 
boundary.  One could obtain a reasonable toy model with an appropriate nuclear
 Hamiltonian.
	
	More generally, one can consider the bouncing mode on top of the 
buckyballs internal modes.  If one considers only bouncing, one obtains a one 
\begin{itshape}dof\end{itshape} Hamiltonian.  This nuclear degree-of-freedom, 
which describes the height of the ball, is coupled to a single electronic 
state.  In this exactly-solvable model, one treats the nucleus as a point 
mass.  The nuclear Hamiltonian consists of the harmonic potential plus a 
second term which contributes nothing when the electron is off the buckyball 
and shifts the oscillator when it is on it.  That is, the nuclear Hamiltonian 
$H_N$ takes the form
\begin{equation}
	H_N = \frac{1}{2}k(z - z_0)^2 + e\mathcal{E}z \hat{n},
\end{equation}
where $k$ is a spring constant, $z$ is the displacement of the buckyball, $e$ 
is the charge of an electron, $\mathcal{E}$ is the electric field, and the 
operator $\hat{n}$ is 1 when the electron is on the buckyball and 0 when it is
 not.  The electronic Hamiltonian is
\begin{equation}
	H_e = E_0 - e\mathcal{E}z,
\end{equation}
where $E_0$ is the energy of the electron without the coupling.  One can then 
complicate matters by modeling the geometry of the buckyball as spherical.  
Internal nuclear modes then give additional nuclear 
\begin{itshape}dof\end{itshape}, and their coupling to multiple electronic 
states is also relevant.  That is, when considering the internal modes of the 
buckyball, it becomes important to consider multiple electronic states of the 
molecule rather than just one.  When treating the buckyball as a point mass, 
it was sufficient to consider only the ground state in the analysis of its 
because the excited states are not energetically close.  Nevertheless, a study
 considering multiple-term superposition states of a zero-dimensional bouncing
 ball might still be illuminating.  When the buckyball is three-dimensional 
rather than zero-dimensional, its (nuclear) deformation can lead smaller 
separations in the eigenenergies, thereby forcing one to consider Gal\"erkin 
projections of more than one electronic state.  However, the number of states 
that should be considered is open to debate.  To answer this question, one 
must consider the buckyball's electronic spectrum.  Moreover, the appropriate 
number of electronic states will depend on the internal (Jahn-Teller) 
distortions under consideration.  One can complicate matters further by 
considering the buckyball's true soccerball geometry rather than a spherical 
approximation.  Unlike a stationary spherical quantum billiard, a billiard 
shaped like a soccerball is no longer globally separable, which leads to a 
marked increase in the complexity of the system's internal dynamics.\cite{nec}
  It is important to note that these internal Jahn-Teller modes can be studied
 without the bouncing dynamics--which essentially become an extra 
degree-of-freedom.  A buckyball is expected to exhibit nonadiabatic behavior 
even without the bouncing mode.  

\begin{centering}
\subsection{Nanotubes}
\end{centering}

	Carbon nanotubes have been studied extensively during the past decade
 from both scientific and engineering perspectives, as they offer numerous 
potential technological applications.\cite{tube}  These compounds may be 
described as cylinders with very large aspect ratios.  That is, their length
 is much larger than their cross-sectional radius, so there are some respects 
in which they can be viewed as one-dimensional objects.  However, nanotubes 
also have hemispherical caps, so they are not truly cylindrical objects 
(although they possess the same reflection and rotation symmetries).  For now,
 we ignore this and treat them as cylinders.  We will briefly revisit this 
issue later and discuss the utility of studying carbon nanotubes using the 
more accurate ``pillbox'' geometry.  Carbon nanotubes may be studied as 
quantum billiards because, like quantum dots, they exhibit ballistic electron 
transport.\cite{marcus2,ball1,ball2}  In particular, there is evidence that 
electrons may traverse the length of some single-walled carbon nanotubes 
ballistically without significant scattering.  Such devices may thus be 
thought of as resonant cavities for electrons in which the nanotube acts as a 
waveguide.  Additionally, the contacts between the nanotube and the electrodes
 act as weakly reflecting barriers.\cite{ball1}

	Under the assumption of a cylindrical geometry, every nanotube is
specified by its diameter and the chirality of the rows of carbon atoms 
relative to the axis of the cylinder.  Their lengths range from the micrometer
 to milimeter scales, and the length of a given nanotube may undergo 
oscillations.  Nanotube diameters, which may also vibrate, range from about 
0.7 to 1.6 nm.  (In contrast, buckyballs have a diameter of about 0.72 nm.)  
Because of these radial and longitudinal vibrations and the relevance of the 
semiquantal regime to the present situation, carbon nanotubes are potentially 
describable as two \begin{itshape}dov\end{itshape} vibrating quantum 
billiards.\cite{nec,rect}  That is, they have two nuclear degrees-of-freedom 
that couple to electronic motion.  This description of nanotubes has not yet 
been studied, and it may prove beneficial to do so.  Another vibrational mode 
resembles distortions reminiscent of plucking a guitar string in which the 
entire nanotube oscillates back and forth while retaining its shape.

	Another possible model of carbon nanotubes is a quantum billiard 
shaped like a pillbox, whose geometry consists of a cylinder adjoined to two 
spherical caps.  Thus, unlike the cylindrical quantum billiard, this system is
 not globally separable.\cite{nec}  Consequently, even a stationary pillbox 
quantum billiard, whose longitudinal cross section is a stadium quantum 
billiard\cite{mac}, experiences quantum signatures of classical chaos 
(``quantum chaology''), much like the quantum Sinai 
billiard.\cite{katok,gutz,atomic,ksu}  Therefore, there is a tangible 
difference between this model and the cylindrical quantum billiard, as the 
latter is integrable unless the boundary oscillates.  Analogous to the 
situation with buckyballs, one can treat the pillbox geometry as a 
perturbation of the cylindrical one.

	One aspect of nanotube dynamics that has been studied is nonlinear 
resonance effects and their relation to positional instability.\cite{tube}  
Such resonances provide evidence that the upper and lower limits of nanotube 
diameters may be affected significantly by the system's internal dynamics.  
Sufficiently chaotic motion during the attempted formation of a nanotube could
 preclude organization into such a well-formed structure.  As with other 
polyatomic molecules, one may study the nonlinear dynamics of carbon nanotubes
 using the classical trajectory method.  It has been shown that onset of 
unstable motion occurs rapidly when certain vibrational modes are coupled.  
This large-amplitude motion is caused by low-order nonlinear resonances.  The 
dynamics of the nanotube were observed to depend sensitively on the length of 
their diameters.  It was also observed that the dependence of the positional 
stability on the diameter was correlated with the length of the nanotube, 
thereby implying a dependence of the onset of large-amplitude motion on the 
aspect ratio of the device.  As in many other fields of 
science\cite{powerlaw}, these dependencies obey scaling laws.  

	Positional instability may arise from strong coupling between 
longitudinal and radial (``ring-breathing'') modes.  Sumpter and 
Noid\cite{tube} found that the two modes had a 1:2 frequency ratio, which is 
sometimes called Fermi resonance and is one of the stronger forms of nonlinear
 coupling.  Such resonances have been attributed to the onset of readily 
manifested energy transfer in several polyatomic and macromolecules.  In the 
present situation, rapid energy transfer between low-frequency modes (which 
retain their energy because of resonant transfer between longitudinal and 
radial motion) causes small-diameter carbon nanotubes to become unstable in 
the sense that they exhibit large-amplitude motion.  Additionally, nanotubes 
are expected to exhibit Jahn-Teller distortions that could significantly 
affect their electronic structure,\cite{full} so a nonadiabatic 
semiquantal description should be appropriate to study the dynamics of these 
devices.  Quantum billiards may also be useful to study other devices such as 
the horn-shaped nanobugles (that are also carbon-based) and the silicon-based 
nanocages that can be formed surrounding a metal ion.
	
\begin{centering}
\section{Other Applications}
\end{centering}

	There are numerous other systems in which semiquantum chaos occurs.  
For example, consider a collection of atoms in a resonant cavity interacting 
self-consistently with the electromagnetic field within the 
cavity.\cite{vibline}  Berman, Bulgakov, and Zaslavsky\cite{berman} studied 
this system \begin{itshape}quasiclassically\end{itshape} to show that the 
quasiclassical approximation may break down in a shorter time frame for a 
system whose classical limit ($\hbar = 0$) is chaotic than it does for one 
whose limit is integrable.  In the quasiclassical regime, the quantity
\begin{equation}
	\frac{\hbar}{I} \ll 1.
\end{equation}
(The variable $I$ is a characteristic action of the problem.)  For integrable 
systems, the approximation breaks down for time $\tau_\hbar$ given by
\begin{equation}
	\tau_\hbar = C/\hbar,
\end{equation}
where $C$ is a constant.  For chaotic systems, however, the breakdown time is 
instead about
\begin{equation}
	\tau_\hbar = C \ln \left(\frac{C}{\hbar}\right).
\end{equation}
Ideally, one studies the dynamics of atoms in a resonant cavity by analyzing 
two electronic states (i.e., using a two-mode Gal\"erkin expansion).  Suppose 
that the electromagnetic field has a single mode whose frequency equals the 
transition frequency of the atoms.  (This situation is encompassed by a model 
attributed to Dicke.)  This system was approximated quasiclassically using a 
\begin{itshape}resonance approximation\end{itshape}, which provides a basis 
for understanding the dynamics of many problems involving the interaction of 
atoms and fields.  (One can also treat this system semiquantally.)  In their 
analysis, Berman, Bulgakov, and Zaslavsky constructed a theory to obtain the 
quantum corrections for the dynamics of this resonant cavity system.  In 
particular, they derived equations of motion for the quantum-mechanical 
expectation values and quantum correlation functions. They demonstrated 
numerically that the nature of the growth of the quantum corrections depended 
very strongly on whether the classical limit was integrable or chaotic.  This 
underscores a very important aspect of quantum chaos: one obtains systems that
 behave in a fundamentally different manner depending on whether one has 
quantized a classically chaotic or classically integrable system.

	Another situation in which the semiquantal approximation is used to 
analyze nonadiabatic dynamics is in collective nuclear motion.\cite{bulgac}  
Near level-crossings, in which the relative energy between two eigenstates 
changes sign, there are several effects that can occur during the time 
evolution of the slow (nuclear) variables.  Among them are Landau-Zener 
transitions, the molecular Aharanov-Bohm effect, and geometric phase, a 
non-integrable quantum phase that we discussed previously.  Moreover, the 
concepts of quantum chaos and level crossings go hand in hand, which the 
reader may have already gathered from prior discussions in this review.  
Traditional studies of large-amplitude collective motion do not discuss 
phenomena such as Berry phase, despite the fact that microscopic computable 
quantities such as potential energies vary rapidly (or may even be singular or
 nearly so) when sufficiently close to level crossings.  Bulgac\cite{bulgac} 
analyzed a simple model to illustrate the effects of level crossings on bound 
nuclear collective motion.

	Another relevant system is a quantum spin $\vec{\sigma}$ interacting 
with the motion of a particle.\cite{real}  The Hamiltonian of one such system 
is
\begin{equation}
	H = B\sigma_z + Cx\sigma_x + \frac{p^2}{2m} + V(x),
\end{equation}
where the first term is the spin Hamiltonian, the last two terms are the 
particle Hamiltonian, and the second term represents the interaction between 
particle and spin.  This system has been explored in both the fully quantum 
and semiquantum regimes.

	Nonadiabatic coupling in semiquantal physics is also relevant to the 
study of micromasers.\cite{haroche}  Consider a single two-level atom in a 
cavity that is coupled to a ``high $Q$'' cavity mode.  (High $Q$ regimes are 
ones in which the cavity has a small number of modes interacting with the 
atomic system.)  This regime is relevant, for example, to superconducting 
microwave cavities and extremely high finesse optical resonators.)  Suppose 
that an atom, initially in an excited state $e$, interacts with an initially 
empty cavity for a time $T$ before exiting.  Moreover, suppose that another 
excited electron enters the cavity before the resultant field from the 
previous electron has had time to relax.  This second electron is consequently
 affected both by the field and the cavity.  Suppose that this process 
continues \begin{itshape}ad infinitum.\end{itshape}  In a steady-state 
operation, the field in the cavity results from competition between relaxation
 and the interaction with successive electrons.  The system so obtained is a 
\begin{itshape}micromaser,\end{itshape} which is easily seen to be amenable to
 a semiquantal description.  

	Finally, consider radio frequency superconducting 
quantum-interference-device (SQUID) magnetometers.\cite{squid}  Such systems
 consist of a superconducting weak link ring and an $LC$ oscillator circuit (a
 tank circuit), which is driven by an external source of current (ordinarily 
at radio frequencies).  The current leads to a magnetic flux in the inductor, 
which interacts with the SQUID ring via a mutual inductance $M$.  This coupled
 magnetic flux induces a screening current in the ring that is also coupled to
 the tank circuit.  This model exhibits chaotic dynamics, which has also been 
reported in experimental rf-SQUID systems.  To incorporate quantum-mechanical 
information into this model, one quantizes the motion of the ring.  One 
thereby obtains a semiquantal system with a quantum component (the ring) 
coupled to a linear classical oscillator (the tank circuit).  This system has 
been shown to exhibit semiquantum chaos.

\begin{centering}
\section{Conclusions}
\end{centering}

	Every physical regime is an approximation of reality in some form or 
another.  One lesser-known regime is the semiquantal one, which may be used to
 describe systems with both classical and quantum subcomponents.  In the 
present review, we discussed nonadiabatic dynamics in the semiquantal regime. 
 We focussed on the arena of electronic-nuclear coupling in molecular 
dynamics, but we also included examples from several other situations.  We 
formulated the notion of semiquantal physics and then discussed nonadiabatic 
phenomena, concentrating on their relation to the Born-Oppenheimer 
approximation.  We also discussed several systems in which such behavior can 
occur. 

\vspace{-.05 in}

\begin{centering}
\section{Acknowledgements}
\end{centering}

	I would like to thank Greg Ezra for carefully reading an early version
 of this manuscript, helping me correct several mistakes, and improving my 
understanding of physical chemistry with explanations and references to 
numerous journal articles.  I would also like to than Paul McEuen for useful 
discussions during the preparation of this manuscript.  Additionally, Greg 
Colyer alerted me to the study of chaotic dynamics in SQUIDs.  Finally, the 
referees made several excellent suggestions that improved the exposition in 
this paper.

\begin{centering}
\bibliographystyle{phaip}
\bibliography{ref}
\end{centering}

\end{document}